\documentclass[9pt,twocolumn,twoside]{osajnl}

\journal{ol} 

\setboolean{shortarticle}{true}

\usepackage{lineno}

\title{Novel phase unwrapping approach based on lateral resolution}

\author[1]{Al\.{ı}m Yolalmaz}
\author[1]{Jeroen Kalkman}

\affil[1]{Department of Imaging Physics, Delft University of Technology, Lorentzweg 1, 2628 CJ Delft, The Netherlands}

\affil[*]{Corresponding author: A.Yolalmaz@tudelft.nl}

\begin{abstract}
The phase unwrapping plays a key role in obtaining a ground-truth phase of the wrapped phase. High-accurate unwrapped phases are demanded in various research fields such as optical holography, optical diffraction tomography, and magnetic resonance imaging. Unfortunately, the ground-truth phase is not accessible due to 2$\pi$ ambiguity which arises from phase jumps in the wrapped phase. In this paper, we propose a novel approach to improve the accuracy of unwrapping process. We increase the sampling frequency by employing a magnifying objective to reconstruct the unwrapped phase with high accuracy for the application of optical holography. Our result shows that optical magnification enables us to improve the accuracy of the true phase by 42\%. We deeply believe that our approach will demonstrate significant achievement in obtaining ground-truth phases in various research fields such as optical holography, magnetic resonance imaging, and optical diffraction tomography.

\end{abstract}

\begin{document}

\maketitle

\section{Introduction}

Phase unwrapping is a crucial process employed in various research fields such as optical interferometry\cite{Ahmad2001, Davidson1999}, seismology\cite{Wang2000}, signal processing\cite{Guerriero1998}, fringe projection profilometry\cite{Wang2023}, magnetic resonance imaging\cite{Strand1999, Langley2009}, and optical diffraction tomography\cite{Horst2020}. Unwrapping a phase is a mandatory process to reach the ground-truth phase from the wrapped (principal) phase. The principal phase is a result of the arctan function experiencing the phase modulo 2$\pi$, and that limits the principal phase in the interval (-$\pi$, $\pi$). Thus, a continuous phase variation of the ground-truth phase is expressed in terms of the discontinuous phase, and this 2$\pi$ ambiguity is realized.

To reach the ground-truth phase and mitigate 2$\pi$ ambiguity, an appropriate integer multiple of 2$\pi$ has to be added to the principal value; thus, this process yields a true (absolute) phase. However, the true phase may not match the ground-truth phase due to various factors such as various noise mechanisms and a high number of phase jumps in the wrapped phase. There are many algorithms and methods to reach the ground-truth phase from the principal phase\cite{Wang1998, Xu2013, Ahmad2001, Guerriero1998, Davidson1999, Carballo2000, Strand1999, Servin1999, Gutmann2000, Xu2022, Xie2021}. These algorithms also show great performance in noisy data\cite{Adi2010, Quiroga1994}. Due to great advancements in technology, deep learning-based unwrapping operations are also available to attain the ground-truth phase with the principal phase\cite{Luo2023, Wang2023}.

In this study, we propose the idea of increasing the sampling frequency of a wrapped phase to reconstruct the ground-truth phase. To improve the sampling frequency, we employ a magnifying objective which allows us to fill in missing information between two consecutive wrapped phase points. Therefore, the unwrapping operation becomes less sensitive to 2$\pi$ ambiguity. As a result of this operation, the true phase becomes less deviated from the ground-truth phase.

\section{Results and discussion}

In Fig. \ref{fig4}a, we present a phase object using the specification of a capillary tube immersed in water. The capillary tube has inner and outer diameters of 0.9 mm and 1.6 mm, respectively. As it is seen in Fig. \ref{fig4}a, the refractive index of the tube is constant over the structure and has a value of 1.344. The surrounding medium is water with a refractive index of 1.333. In  Fig. \ref{fig4}b, clear sharp RI variation is demonstrated along the line at z=0. The corresponding phase shift on an image plane is computed for lensless imaging case with unity magnification by using Eq. \ref{Eq1} and considering the wavelength of a light source $\lambda$ as 633 nm. $\phi(x, y)$, $z$, and $\Delta n$ are the ground-truth phase on the spatial coordinates $x$ and $y$, the propagation distance of light through the capillary tube along the z-axis, and the refractive index difference between the capillary tube and water, respectively. The ground-truth phase distribution is presented in Fig. \ref{fig4}c where due to the cylindrical symmetry of the capillary tube, along the y-axis, the phase shift is not a function of y-axis. In Fig. \ref{fig4}d, phase variation along x-axis is seen due to the presence of two media: water and the capillary tube.

\begin{equation}{
\phi(x, y)=\int \frac{2 \pi}{\lambda} \Delta n(x, y, z) d z}
\label{Eq1}
\end{equation}

The electric field that describes a phase object is analytically formulated by Eq. \ref{Eq2}, where $U(x, y)$ is the electric field of the light. When this electric field is indirectly recorded by a recording medium such as a CCD camera, its corresponding phase distribution is acquired after computing the argument of the complex electric field by using Eq. \ref{Eq3} where $\phi_p(x, y)$ is the principal phase distribution of the phase object. As a result of this operation, the principal phase of the phase object is realized as seen in Fig. \ref{fig4}e. Unfortunately, due to the arctan operation, the principal phase values are restricted in the interval (-$\pi$, $\pi$). This is clearly observed in Fig. \ref{fig4}f when we look at the phase variation along the line y=0 mm in Fig. \ref{fig4}e.

\begin{equation}{
U(x, y)=\mathrm{e}^{\mathrm{j} \phi(x, y)}}
\label{Eq2}
\end{equation}

\begin{equation}{\phi_p(x, y)=\operatorname{Arg}[U(x, y)]}
\label{Eq3}
\end{equation}

\begin{equation}{
\phi_t(x, y)=\phi_p(x, y)+2 \pi m(x, y)}
\label{fig4}
\end{equation}

 \begin{figure}[!htb]
    \centering
        \includegraphics[width=82 mm]{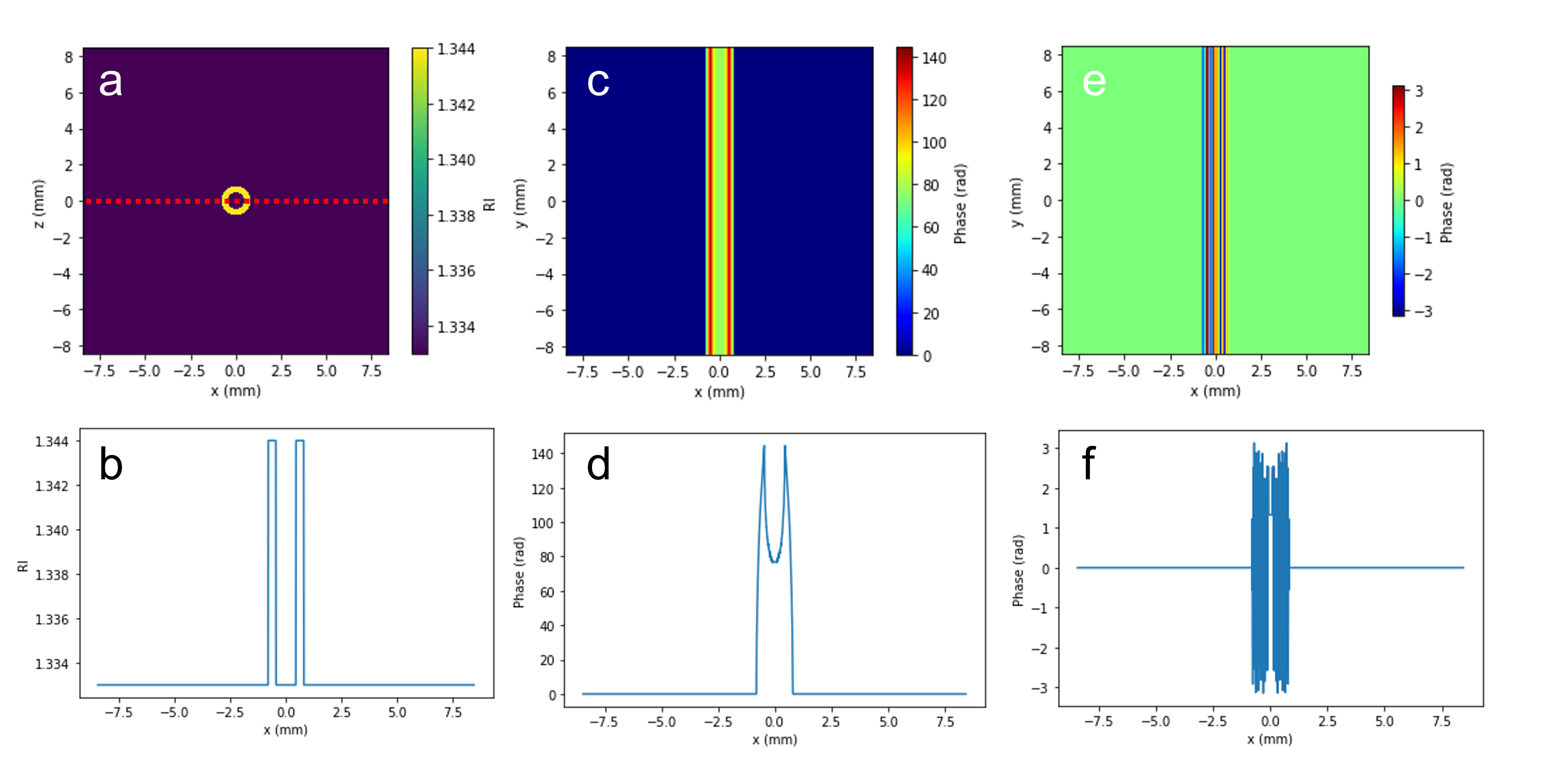}
        \caption{(a) RI distribution of a capillary tube, (b) RI variation along the red line in (a), (c) Ground-truth phase distribution of the capillary tube, (d) Phase distribution of the capillary tube along y = 0 in (c), (e) Principal phase distribution of the capillary tube in (c), (f) Cross-sectional phase variation along y = 0 in (e). }
    \label{fig4}
\end{figure}

\begin{figure}[!htb]
    \centering
        \includegraphics[width=82 mm]{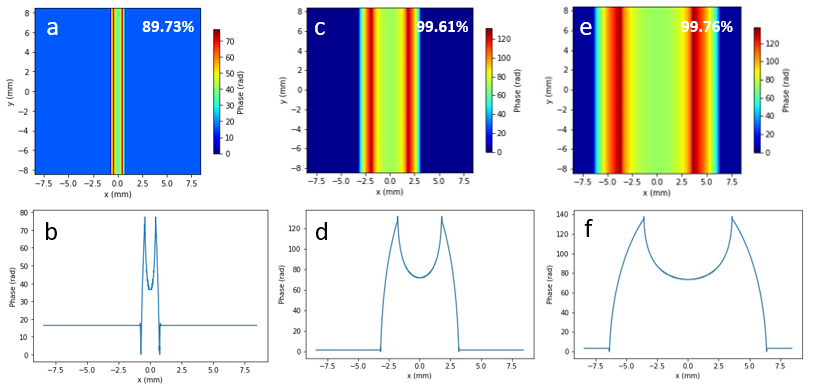}
        \caption{True phase distributions as a result of unwrapping operations. (a) Before magnifying the capillary tube using an objective, (b) cross-sectional phase variation along y = 0 in (a), (c) after magnification of the capillary tube by a 4 times magnifying objective, (d) cross-sectional phase variation along y = 0 in (c), (e) after magnification of the capillary tube by an 8 times magnifying objective, (f) cross-sectional phase variation  along y = 0 in (e).}
    \label{fig2}
\end{figure}

The true phase of the phase object is recovered by the sorting-based phase unwrapping algorithm in this study\cite{Herraez2002}. As a result of the unwrapping process, we obtain a true phase distribution (see Fig. \ref{fig2}a). We observe a similar trend in phase variation along the y-axis in Fig. \ref{fig2}b, but the maximum phase shift of 77.3 rad does not match with the ground-truth value of 144.5 rad. The reason behind the difference is the high number of phase jumps in the principal phase distribution (see Fig. \ref{fig4}f); thus, the unwrapping algorithm yields the incorrect true phase, and we encounter an accuracy of 53.5 \%. When this phase distribution is compared to the ground-truth phase (Fig. \ref{fig4}c), Pearson's correlation coefficient becomes 0.8973 meaning 89.73\% similarity of the true phase with respect to the ground-truth phase. Considering the importance of phase recovery and thickness measurement, this error leads to a 615 $\mu$m deviation of thickness from the ground-truth thickness with the simulation parameters. This thickness difference is too high and severe, especially in imaging microscopic structures.

\begin{figure}[!htb]
    \centering
        \includegraphics[width=82 mm]{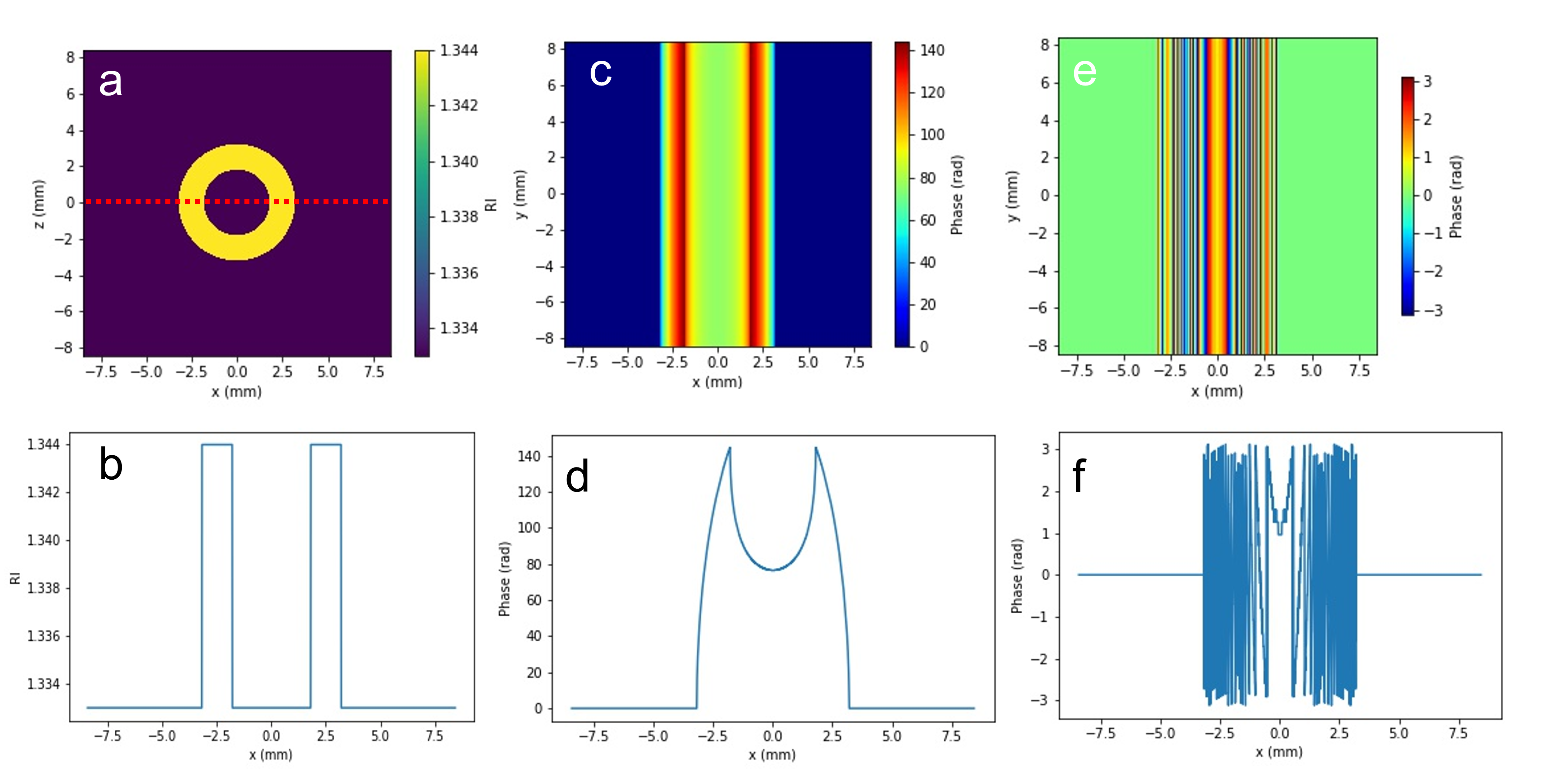}
        \caption{After magnifying the capillary tube by an objective of 4 times magnification factor. (a) RI distribution of a capillary tube, (b) RI variation along the red line in (a), (c) Ground-truth phase distribution of the capillary tube, (d) Phase distribution of the capillary tube along y = 0 in (d), (e) Wrapped phase distribution of the capillary tube in (c), (f) Cross-sectional phase variation along y = 0 in (e). }
    \label{fig1}
\end{figure}

To obtain the ground-truth phase and minimize this huge inaccuracy in the unwrapping process, in this study, we magnify the capillary tube by an objective with a 4 times magnification factor. In Fig. \ref{fig1}a, the refractive index distribution of the tube within the water is presented. The spatial distance between the inner and outer circumferences of the capillary tube is hypothetically increased as seen in the refractive index distribution of the phase object (Fig. \ref{fig1}b). That leads us to express the same phase variation with a higher number of data points compared to the case where the object is not magnified. As a result of optical magnification, we disperse the lateral phase distribution of the phase object over a larger area on the image plane in Fig. \ref{fig1}c. When cross-sectional phase variations in Fig. \ref{fig4}d and Fig. \ref{fig1}d are crosschecked, we come up with a higher number of data points expressing the same phase shift between the capillary tube and water. After computing Eq. \ref{Eq2} and Eq. \ref{Eq3} sequentially with the phase distribution in Fig. \ref{fig1}c, we reach the principal phase of the capillary tube magnified by 4 times in Fig. \ref{fig1}e. As a result of the magnification, the principal phase demonstrates a higher number of phase jumps compared to the non-magnified situation (see Fig. \ref{fig4}f and Fig. \ref{fig1}f). Furthermore, the number of data points between two consecutive peak points is enhanced (see Fig. \ref{fig1}f).  

The true phase of the phase object after the 4 times magnification is retrieved in Fig. \ref{fig2}c by implementing the same unwrapping algorithm for the principal phase in Fig. \ref{fig1}e. When this phase distribution is compared to the ground-truth phase (Fig. \ref{fig1}c), Pearson's correlation coefficient becomes 99.6\%. The maximum phase shift is 131.5 rad in the phase distribution which has an accuracy of 91\%  when compared to the ground-truth phase shift of 144.5 rad (Fig. \ref{fig1}c). The unwrapping process produces an accuracy improvement from 53.5\% to 91.0\% which is a significant achievement after the phase object is magnified. Considering our results, we are subject to fewer 2$\pi$ ambiguities in the unwrapping process due to higher sampling of the phase object thanks to the magnifying objective.

Here we also demonstrate the results of magnification with an 8 times magnifying objective. It is observed in Fig. \ref{fig2}e where the phase shift between the capillary tube and water is densely dispersed within the image plane after magnifying the phase object and unwrapping operation. The true phase shows a 99.8\% correlation with respect to the ground-truth phase distribution. Moreover, the true phase of the object has a maximum of 137.7 rad phase shift which is 95.3\% accurate (see Fig. \ref{fig2}f).

\section{Conclusion}

In this work, we obtain a significant accuracy improvement in retrieving the true phase in the unwrapping operation. Our novel approach, which is based on the lateral magnification of the phase object, increases the sampling frequency and provides highly accurate phase reconstruction. We believe that our idea is benefited from various research fields such as optical holography to obtain 3D object information with better accuracy.

\section{acknowledgments}
The authors are grateful for the financial support from Medical Delta Consortium.

\bibliography{Library_Unwrapping.bib}

\begin{thebibliography}{10}
\newcommand{\enquote}[1]{``#1''}

\bibitem{Ahmad2001}
F.~H. Ahmad, G.~L. Helms, R.~M. Castellane, and B.~P. Durst, {\protect\JournalTitle{Microwave and Optical Technology Letters}} \textbf{32}, 101 (2001).

\bibitem{Davidson1999}
G.~Davidson and R.~Bamler, {\protect\JournalTitle{{IEEE} Transactions on Geoscience and Remote Sensing}} \textbf{37}, 163 (1999).

\bibitem{Wang2000}
Y.~Wang, {\protect\JournalTitle{Journal of Seismic Exploration}} \textbf{9}, 93 (2000).

\bibitem{Guerriero1998}
L.~Guerriero, G.~Nico, G.~Pasquariello, and S.~Stramaglia, {\protect\JournalTitle{Applied Optics}} \textbf{37}, 3053 (1998).

\bibitem{Wang2023}
S.~Wang, T.~Chen, M.~Shi, D.~Zhu, and J.~Wang, {\protect\JournalTitle{Optics and Lasers in Engineering}} \textbf{162}, 107409 (2023).

\bibitem{Strand1999}
J.~Strand and T.~Taxt, {\protect\JournalTitle{Applied Optics}} \textbf{38}, 4333 (1999).

\bibitem{Langley2009}
J.~Langley and Q.~Zhao, {\protect\JournalTitle{Magnetic Resonance Imaging}} \textbf{27}, 1293 (2009).

\bibitem{Horst2020}
J.~van~der Horst, A.~K. Trull, and J.~Kalkman, {\protect\JournalTitle{Optica}} \textbf{7}, 1682 (2020).

\bibitem{Wang1998}
Y.~Wang, {\protect\JournalTitle{Journal of Seismic Exploration}} \textbf{7}, 109 (1998).

\bibitem{Xu2013}
Z.~Xu, B.~Huang, and S.~Xu, {\protect\JournalTitle{Electronics Letters}} \textbf{49}, 1565 (2013).

\bibitem{Carballo2000}
G.~Carballo and P.~Fieguth, {\protect\JournalTitle{{IEEE} Transactions on Geoscience and Remote Sensing}} \textbf{38}, 2192 (2000).

\bibitem{Servin1999}
M.~Servin, F.~J. Cuevas, D.~Malacara, J.~L. Marroquin, and R.~Rodriguez-Vera, {\protect\JournalTitle{Applied Optics}} \textbf{38}, 1934 (1999).

\bibitem{Gutmann2000}
B.~Gutmann and H.~Weber, {\protect\JournalTitle{Applied Optics}} \textbf{39}, 4802 (2000).

\bibitem{Xu2022}
C.~Xu, Y.~Cao, H.~Wu, H.~Li, H.~Zhang, and H.~An, {\protect\JournalTitle{Optical Engineering}} \textbf{61}, 4 (2022).

\bibitem{Xie2021}
X.~Xie and Q.~zeng, {\protect\JournalTitle{Optics and Lasers in Engineering}} \textbf{142}, 106615 (2021).

\bibitem{Adi2010}
K.~Adi, A.~B. Suksmono, T.~L.~R. Mengko, and H.~Gunawan, {\protect\JournalTitle{{IEEE} Geoscience and Remote Sensing Letters}} \textbf{7}, 704 (2010).

\bibitem{Quiroga1994}
J.~A. Quiroga and E.~Bernabeu, {\protect\JournalTitle{Applied Optics}} \textbf{33}, 6725 (1994).

\bibitem{Luo2023}
X.~Luo, W.~Song, S.~Bai, Y.~Li, and Z.~Zhao, {\protect\JournalTitle{Optics {\&} Laser Technology}} \textbf{163}, 109340 (2023).

\bibitem{Herraez2002}
M.~A. Herr{\'{a}}ez, D.~R. Burton, M.~J. Lalor, and M.~A. Gdeisat, {\protect\JournalTitle{Applied Optics}} \textbf{41}, 7437 (2002).

\end{thebibliography}

\end{document}